\documentclass[
aps,superscriptaddress,
twocolumn,prd,
showpacs,preprintnumbers,nofootinbib,
amsmath,amssymb,
aps,
superscriptaddress]{revtex4-1}
\usepackage[normalem]{ulem}
\usepackage{graphicx}
\usepackage{soul}
\usepackage{amssymb}
\makeatletter
\let\saved@includegraphics\includegraphics
\AtBeginDocument{\let\includegraphics\saved@includegraphics}
\renewenvironment*{figure}{\@float{figure}}{\end@float}
\usepackage{color}
\usepackage[dvipsnames]{xcolor}
\usepackage{amsmath}
\usepackage{adjustbox}
\makeatletter

\usepackage{tabularx}
\usepackage{hyperref}

\newcommand{\etk}{\textsc{Einstein Toolkit }}
\newcommand{\carpet}{\textsc{Carpet }}

\begin{document}


\title{GW190521 as a merger of Proca stars: \\ a potential new vector boson of $8.7 \times 10^{-13}$ eV}


\author{Juan Calder\'on~Bustillo}
 	\email{juan.calderon.bustillo@gmail.com}
	\affiliation{Instituto Galego de F\'{i}sica de Altas Enerx\'{i}as, Universidade de
Santiago de Compostela, 15782 Santiago de Compostela, Galicia, Spain}
	\affiliation{Department of Physics, The Chinese University of Hong Kong, Shatin, N.T., Hong Kong}
	\affiliation{Monash Centre for Astrophysics, School of Physics and Astronomy, Monash University, VIC 3800, Australia}
	\affiliation{OzGrav: The ARC Centre of Excellence for Gravitational-Wave Discovery, Clayton, VIC 3800, Australia}
\author{Nicolas Sanchis-Gual}
	\email{nicolas.sanchis@tecnico.ulisboa.pt}
	\affiliation{Centro de Astrof\'{i}sica e Gravita\c{c}\~{a}o - CENTRA,
Departamento de F\'{i}sica, Instituto Superior T\'{e}cnico - IST, Universidade de Lisboa - UL, Avenida Rovisco Pais 1, 1049-001, Portugal}
	\affiliation{Departamento  de  Matem\'{a}tica  da  Universidade  de  Aveiro  and  Centre  for  Research  and  Development in  Mathematics  and  Applications  (CIDMA),  Campus  de  Santiago,  3810-183  Aveiro,  Portugal}
\author{Alejandro Torres-Forn\'e}
	\affiliation{Max Planck Institute for Gravitational Physics (Albert Einstein Institute), Am M\"uhlenberg 1, Potsdam 14476, Germany}
         \affiliation{Departamento de Astronom\'{i}a y Astrof\'{i}sica, Universitat de Val\`{e}ncia,
Dr. Moliner 50, 46100, Burjassot (Val\`{e}ncia), Spain}
	\affiliation{Observatori Astron\`{o}mic, Universitat de Val\`{e}ncia,
C/ Catedr\'{a}tico Jos\'{e} Beltr\'{a}n 2, 46980, Paterna (Val\`{e}ncia), Spain}
\author{Jos\'e A. Font}
	\affiliation{Departamento de Astronom\'{i}a y Astrof\'{i}sica, Universitat de Val\`{e}ncia,
Dr. Moliner 50, 46100, Burjassot (Val\`{e}ncia), Spain}
	\affiliation{Observatori Astron\`{o}mic, Universitat de Val\`{e}ncia,
C/ Catedr\'{a}tico Jos\'{e} Beltr\'{a}n 2, 46980, Paterna (Val\`{e}ncia), Spain}
\author{Avi Vajpeyi}
	\affiliation{Monash Centre for Astrophysics, School of Physics and Astronomy, Monash University, VIC 3800, Australia}
	\affiliation{OzGrav: The ARC Centre of Excellence for Gravitational-Wave Discovery, Clayton, VIC 3800, Australia}
\author{Rory Smith}
	\affiliation{Monash Centre for Astrophysics, School of Physics and Astronomy, Monash University, VIC 3800, Australia}
	\affiliation{OzGrav: The ARC Centre of Excellence for Gravitational-Wave Discovery, Clayton, VIC 3800, Australia}
\author{Carlos Herdeiro} 
		\affiliation{Departamento  de  Matem\'{a}tica  da  Universidade  de  Aveiro  and  Centre  for  Research  and  Development in  Mathematics  and  Applications  (CIDMA),  Campus  de  Santiago,  3810-183  Aveiro,  Portugal}
\author{Eugen Radu}
	\affiliation{Departamento  de  Matem\'{a}tica  da  Universidade  de  Aveiro  and  Centre  for  Research  and  Development in  Mathematics  and  Applications  (CIDMA),  Campus  de  Santiago,  3810-183  Aveiro,  Portugal}
\author{Samson H. W. Leong}
	\affiliation{Department of Physics, The Chinese University of Hong Kong, Shatin, N.T., Hong Kong}


\begin{abstract}
Advanced LIGO-Virgo have reported a short gravitational-wave signal (GW190521) interpreted as a quasi-circular merger of black holes, one at least populating the pair-instability supernova gap, that formed a remnant black hole of $M_f\sim 142 M_\odot$ at a luminosity distance of $d_L \sim 5.3$ Gpc.
With barely visible pre-merger emission, however, GW190521 merits further investigation of the pre-merger dynamics and even of the very nature of the colliding objects.
We show that GW190521 is consistent with numerically simulated signals from head-on collisions of two (equal mass and spin) horizonless vector boson stars (aka Proca stars), forming a final black hole with $M_f = 231^{+13}_{-17}\,M_\odot$, located at a distance of $d_L = 571^{+348}_{-181}$~Mpc.
This provides the first demonstration of close degeneracy between these two theoretical models, for a real gravitational-wave event.
The favoured mass for the ultra-light vector boson constituent of the Proca stars is $\mu_{\rm V}= 8.72^{+0.73}_{-0.82}\times10^{-13}$~eV.
 Confirmation of the Proca star interpretation, which we find statistically slightly preferred, would provide the first evidence for a long sought dark matter particle.

\end{abstract}

\maketitle

\begin{figure*}
\begin{center}
\includegraphics[width=0.493\textwidth]{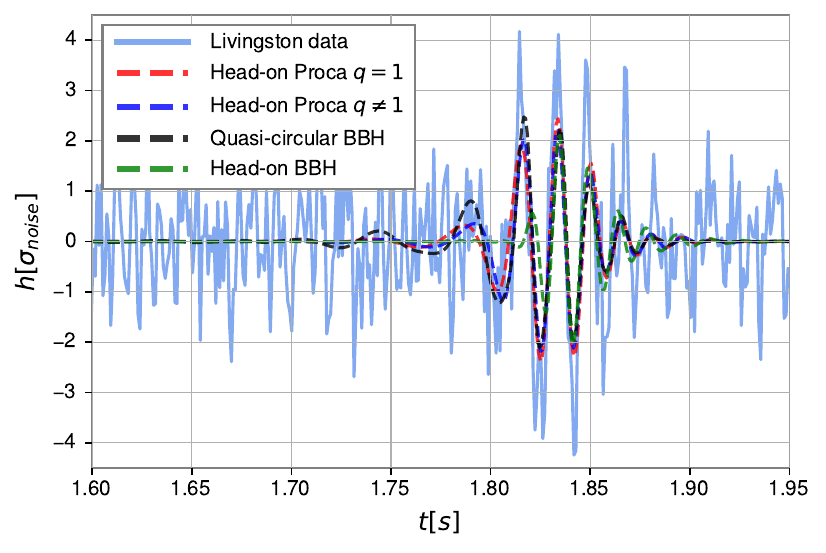}
\includegraphics[width=0.50\textwidth]{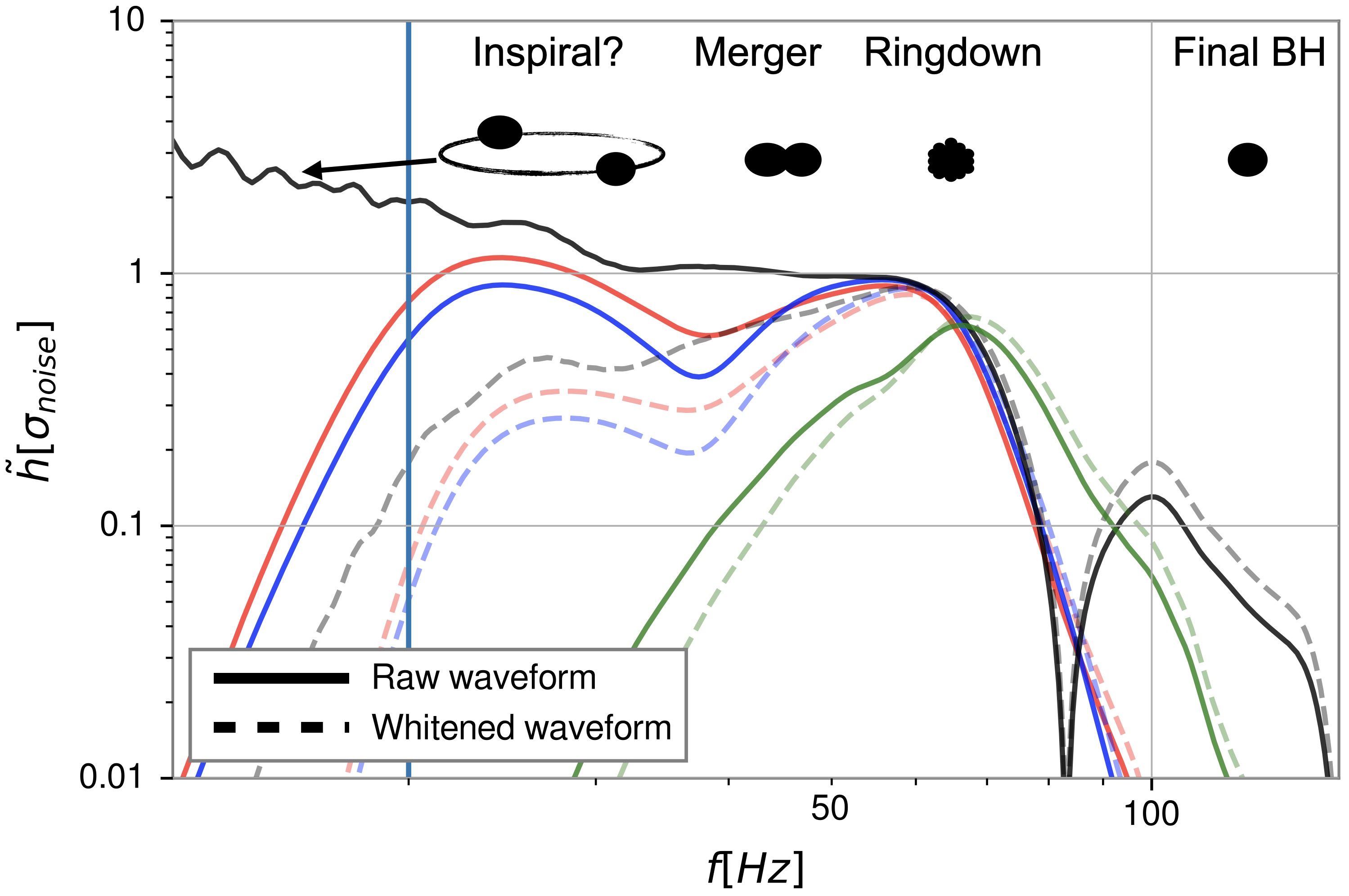}
\caption{\textbf{Time-series and spectrum of GW190521.} \textbf{Left:} Whitened strain data of the LIGO Livingston detector at the time of GW190521, together with the best fitting waveforms for a head-on merger of two BHs (green), two equal/unequal mass PSs (red and blue) and for a quasi-circular BH merger (black). The time axis is expressed so that the GPS time is equal to $t_{\text{GPS}}= t +1242442965.6069$ s. \textbf{Right:} corresponding waveforms shown in the Fourier domain. Solid lines denote raw waveforms (scaled by a suitable, common factor) while dashed lines show the whitened versions. The vertical line denotes the $20$ Hz limit, below which the detector noise increases dramatically. Due to this, a putative inspiral signal from a quasi-circular BBH merger (solid black) would be almost invisible to the detector (see dashed grey) and barely distinguishable from PHOC signals (dashed red and blue).}
\label{fig:strain}
\end{center}
\end{figure*}


 {\it \textbf{Introduction.}} Gravitational-wave (GW) astronomy has revealed stellar-mass black holes (BHs) more massive than those known from X-ray observations \cite{Abbott:2016blz,LIGOScientific:2018mvr}. This population, with masses of tens of solar masses, complements the supermassive black holes (SMBHs) lurking
in the centre of most galaxies, with masses in the range  $10^5-10^{10}\,M_\odot$ \cite{Volonteri2010}. The observation of GW190521 \cite{GW190521D} by the Advanced LIGO \cite{AdvancedLIGOREF} and Virgo \cite{TheVirgo:2014hva} detectors 
has populated the gap between these two extremes. The LIGO-Virgo Collaboration (LVC) interprets GW190521 as a short-duration signal consistent with a quasi-circular binary black hole (BBH) merger, with mild signs of orbital precession, that left 
behind the first ever observed intermediate-mass black hole (IMBH), with a mass of $\sim 142 M_\odot$ \cite{GW190521D,GW190521I}. 
This interpretation is challenged by the fact that at least one of the BHs sourcing GW190521 must fall within the pair-instability supernova (PISN) gap. Alternative interpretations of GW190521 as an eccentric BBH lead to the same conclusion \cite{Isobel_ecc,Gayahtri_ecc}.
According to stellar evolution, such BHs
cannot form from the collapse of a star \cite{Heger:2002by}, suggesting that this event is sourced by second generation BHs, born in previous mergers.

GW190521 is, however, different from
previously observed signals. While consistent with a BBH merger, its pre-merger signal, and therefore a putative inspiral phase, 
is
barely observable in the detectors sensitive band, motivating the exploration of alternative scenarios that do not involve an inspiral stage. One such possibility is a head-on collision (HOC), which we have recently investigated \cite{HeadOnArxiv}. Within such geometry, however, the high spin of the GW190521 remnant, $a\sim 0.7$, is difficult to reach with mass ratios ($1 < q\equiv m_1/m_2 < 4$) due to the lack of orbital angular momentum and the Kerr limit on the BH spin ($a\leq 1$), 
imposed by the cosmic censorship conjecture. There exist, however, exotic compact objects (ECOs) not subject to this limit that may mimic BBH signals, leading to a degeneracy in the emitted signals \cite{PaniCardoso}.

ECOs have been proposed, e.g., as dark-matter candidates, often invoking
the existence of hypothetical ultra-light (i.e.~sub-eV) bosonic particles. One common candidate is the pseudo-scalar QCD axion,
but other ultra-light bosons arise, e.g., in the string axiverse \cite{Arvanitaki2010}. In particular, vector bosons are also motivated in extensions of the Standard Model of elementary particles and can clump together forming macroscopic entities dubbed bosonic
stars. These are amongst the simplest and dynamically more robust ECOs proposed so far and their dynamics has been extensively studied,  $e.g.$ \cite{liebling2017dynamical,bezares2017final,palenzuela2017gravitational,sanchis2019head}. Scalar boson stars and their vector analogues, Proca stars~\cite{brito2016proca,sanchis2017numerical} (PSs), are self-gravitating stationary solutions of the Einstein-(complex, massive) Klein-Gordon~\cite{Schunck:2003kk} and of the Einstein-(complex) Proca~\cite{brito2016proca} systems, respectively. These consist on complex bosonic fields oscillating at a well-defined frequency $\omega$, which determines
the mass and compactness of the star. 
Bosonic stars can dynamically form without any fine-tuned condition through the gravitational cooling mechanism \cite{Seidel1994,DiGiovanni2018}. While spinning solutions have been obtained for both scalar and vector bosons, the former are unstable against non-axisymmetric perturbations \cite{SanchisGual2019}. Hence, we will focus on the vector case in this work. 
For non-self-interacting bosonic fields, the maximum possible mass of the corresponding stars is determined  by the boson particle mass $\mu_V$.  In particular, ultra-light bosons within $10^{-13}\leq\mu_V\leq10^{-10}$~eV, can form stars with maximal masses ranging between $\sim$ 1000 and 1 solar masses, respectively.\\

We perform Bayesian parameter estimation and model selection on 4 seconds of publicly available data \cite{GWOSC} from the two Advanced LIGO and Virgo detectors around the time of GW190521 (for full details see the Supplemental Material, which includes references \cite{Ashton:2018jfp,CPNest,Cornish:2014kda,Cornish2015,Finn1992,Romano2017,Cutler1994} ).
We compare GW190521 to numerical-relativity simulations of (i) HOCs, (ii) equal-mass and equal-spin head-on PS mergers (PHOCs), and (iii) to the surrogate model for generically spinning BBH mergers \texttt{NRSur7dq4} \cite{NRSur7dq4}. Our simulations include the GW modes $(\ell,m) = (2,0),(2, \pm 2),(3,\pm 2)$ while the BBH model contains all modes with $\ell\leq 4$. 
The PHOC cases we consider form a Kerr BH with a feeble Proca remnant that does not impact on the GW emission \cite{sanchis2020synchronised}. Finally, to check the robustness of our results, we perform an exploratory study comparing GW190521 to a limited family of simulations for unequal-mass $(q\neq 1)$ head-on PS mergers. For finer details on numerical simulations, we refer the reader to our Supp. Material and references \cite{EinsteinToolkit,Loffler:2011ay, Zilhao:2013hia,Schnetter:2003rb,Cactus,ansorg2004single,Canuda_2020_3565475, Zilhao:2015tya, Reisswig:2015} therein.\\

{\it \textbf{Results.}}  Figure ~\ref{fig:strain} shows the whitened strain time series from the LIGO Livingston detector and the best fitting waveforms returned by our analyses for HOCs, PHOCs and BBH mergers. While the latter two show a similar morphology with slight pre-peak power, the HOC signal is noticeably shorter and has a slightly larger ringdown frequency.
These features are more evident in the right panel, where we show the corresponding Fourier transforms (dashed) together with the corresponding raw, non-whitened versions (solid). The HOC waveform displays a rapid power decrease at frequencies below its peak due to the absence of an inspiral. In contrast, PHOCs show a low-frequency tail due to the pre-collapse emission that mimics the typical inspiral signal present in the BBH case down to $f\simeq 20$ Hz. Below this limit, the putative inspiral signal from a BBH disappears behind the detector noise (dashed grey) making the signal barely distinguishable from that of a PHOC.

\begin{figure}[t!]
\begin{center}
\includegraphics[width=0.45\textwidth]{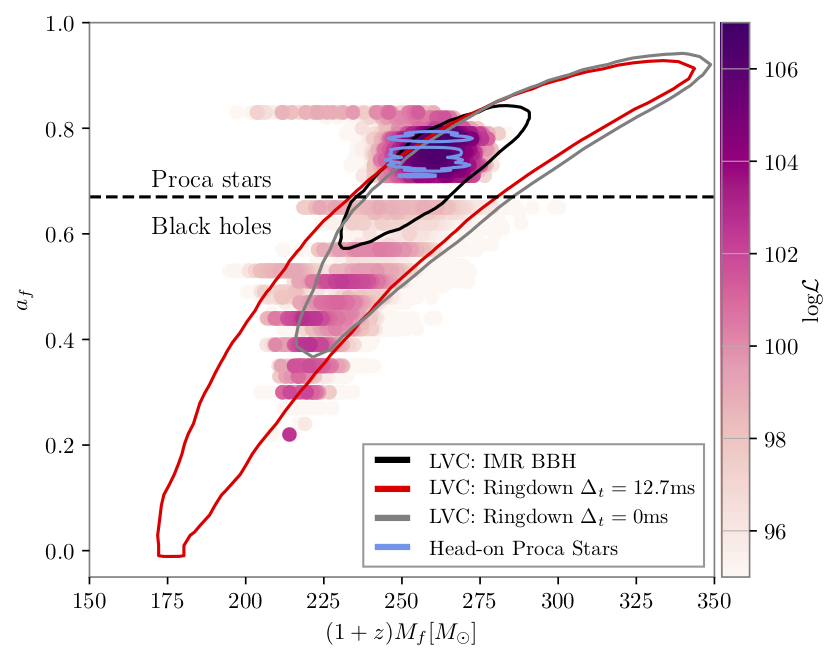}
\caption{
Redshifted final mass and spin of GW190521 according to different waveform models, and directly inferred from a ringdown analysis. The contours delimit $90\%$ credible intervals. For head-on PS and BH mergers (PHOCs and HOCs), we plot the samples colored according to their LogLikelihood. The horizontal dashed line denotes an experimental limit for the final-spin of head-on BH mergers that separates them from head-on PS mergers.}
\label{fig:mass}
\end{center}
\end{figure}

\begin{table}[t!]
\centering
\begin{center}
\begin{tabular}{lcc}
\rule{0pt}{3ex}%
Waveform model & $\log{\cal B}$  &  $\log{\cal L}_{\rm max}$ \\
\hline
\rule{0pt}{3ex}%
Quasi-circular Binary Black Hole & 80.1  & 105.2 \\
\rule{0pt}{3ex}%
Head-on Equal-mass Proca Stars & 80.9  & 106.7\\
\rule{0pt}{3ex}%
Head-on Unequal-mass Proca Stars &  82.0  & 106.5\\
\rule{0pt}{3ex}%
Head-on Binary Black Hole &  75.9 &  103.2 
\\[3pt]

\end{tabular}
\caption{\textbf{Bayesian evidence for our GW190521 source models.} We report the natural Log Bayes Factor obtained for our different waveform models and corresponding maximum values of the Log Likelihood. 
We note that parameter estimation codes \textit{are not} designed to find the true maximum of the likelihood, so that the values we report should be considered as approximate.}
\label{tab:logb1}
\end{center}
\end{table}

Fig. \ref{fig:mass} shows the two-dimensional $90\%$ credible intervals for the redshifted final mass and the final spin obtained by the LVC using BBH models covering inspiral, merger and ringdown (IMR, in black) and solely from the ringdown emission; starting at the signal peak (grey) and 12.7 milliseconds later (pink) \cite{Carullo:2019flw,Isi:2019aib}. Overlaid, we show the red-shifted final mass $M_f^{z}$ and spin $a_f$ obtained by PHOC and HOC models, with the color code denoting the log-likelihood of the corresponding samples. For these, we approximate the final mass by the total mass due to the negligible loss to GWs.

The absence of an inspiral makes HOCs and PHOCs less luminous than BBHs, therefore requiring a lower initial mass to produce the same final BH as a BBH. Accordingly, the BBH scenario yields $M^{z}_{\text{BBH}}=272^{+26}_{-27}\,M_\odot$  \cite{GWOSC,GW190521D} 
while the former two yield lower values of $M^{z}_{\text{HOC}}=238^{+24}_{-21}\,M_\odot$ and $M^{z}_{\text{PHOC}}=258^{+6}_{-8}\,M_\odot$, both consistent within with those estimated by the LVC ringdown analysis, $M^{z}_{\text{BBH, Ringdown}}=252^{+63}_{-64}\,M_\odot$ \cite{GW190521D}, which makes no assumption on the origin of the final BH.

There is, however, a clear separation between HOCs and BBHs/PHOCs in terms of the final spin. Cosmic censorship imposes a bound $a\leq 1$ on the BHs' dimensionless spins  \cite{Wald:1997wa}. This, together with the negligible orbital angular momentum of HOCs, prevents the production a final BH with the large spin predicted by BBH models.
By contrast, PSs are not constrained by $a\leq 1$ and can form remnant BHs with higher spins from head-on collisions. Consequently, the final spin and redshifted mass predicted by PHOCs coincide with those predicted by BBH models. In addition, the discussed lack of pre-peak power in HOCs 
leads to a poor signal fit that penalises the model. In Table~\ref{tab:logb1} we report the Bayesian evidence for our source models. We obtain a relative natural log Bayes factor $\log{\cal{B}}^{\text{HOC}}_{\text{BBH}} \sim - 4.2$ that allows us to confidently discard the HOC scenario.

Unlike BHs, neutron star and PS mergers do not directly form a ringing BH. Instead, a remnant transient object produces GWs before collapsing to a BH, leaving an imprint in the GWs that is not present for HOCs, before emitting the characteristic ringdown signal. For this reason, PHOCs do not only lead to a final mass and spin fully consistent with the LVC BBH analysis but also provide a better fit to the data than HOCs, reflected by a larger maximum likelihood in Table ~\ref{tab:logb1}.\\ 

While BBHs lose around $7\%$ of their mass to GWs, head-on mergers radiate only $\sim 0.1\%$ of it, leading to much lower distance estimates, and consequently, to much larger source-frame masses. 
Whereas the LVC reports a luminosity distance of
 $d_L\sim 5.3^{+2.4}_{-2.6}$ Gpc  \cite{GW190521D}, 
our PHOCs scenario yields
$d_L = 571^{+348}_{-181}$ Mpc, similar to GW150914 \cite{Abbott:2016blz}. Consequently, we estimate a source-frame final mass of $\sim 231^{+13}_{-17}\,M_\odot$, $62\%$ larger than the $142^{+28}_{-16}\, M_\odot$ reported by the LVC. The lower distance estimate handicaps the PHOC model with respect to the BBH one if an uniform distribution of sources in the Universe is assumed. Nonetheless,
Table~\ref{tab:logb1} reports a $\log{\cal{B}}^{\text{PHOC}}_{\text{BBH}} \sim 0.8$, slightly favouring the PHOC model. Relaxing this assumption leads to an increased $\log{\cal{B}}^{\text{PHOC}}_{\text{BBH}} \sim 3.4$ (see Supplementary Material Table I for further details when using this alternative prior). The evidence for the PHOC model is accompanied by a better fit to the data. In addition, BBHs span a significantly larger parameter space that may penalise this model.
While we explored several simplifications of the BBH model (see Supplementary Material), no statistical preference for the BBH scenario was obtained. We therefore conclude that, however exotic, the PHOC scenario is slightly preferred despite being intrinsically disfavoured by our standard source-distribution prior. 

Unlike BBH signals \cite{maggiore2008gravitational}, head-on ones are not dominated by the quadrupole $(\ell,m)=(2,\pm2)$ modes but have a co-dominant $(2,0)$ mode \cite{anninos1995head,palenzuela2007head}. By repeating our analysis removing the $(2,0)$ from our waveforms, we obtain $\log{\cal{B}}^{(2,0)}_{\text{No} (2,0)}=0.6$ in favour of its presence in the signal. The asymmetries introduced by this mode also allow us to constrain the azimuthal angle $\varphi$ describing the projection of the line-of-sight onto the collision plane, normal to the final spin. We estimate $\varphi = 0.65^{+0.86}_{-0.54}$ rad measured from the collision axis, in the direction of any of the two spins (see Supplementary Material, which includes references \cite{Graff:2015bba,CalderonBustillo:2018zuq,CalderonBustillo:2019wwe,Chatziioannou:2019dsz,LIGOScientific:2020stg}). This is, we restrict $\varphi$ to the first and third quadrant of the collision plane, towards where the trajectories of both stars are curved due to frame-dragging. To the best of our knowledge, this is the first time such measurement is performed. 

\begin{figure}
\begin{center}
\includegraphics[width=0.22\textwidth]{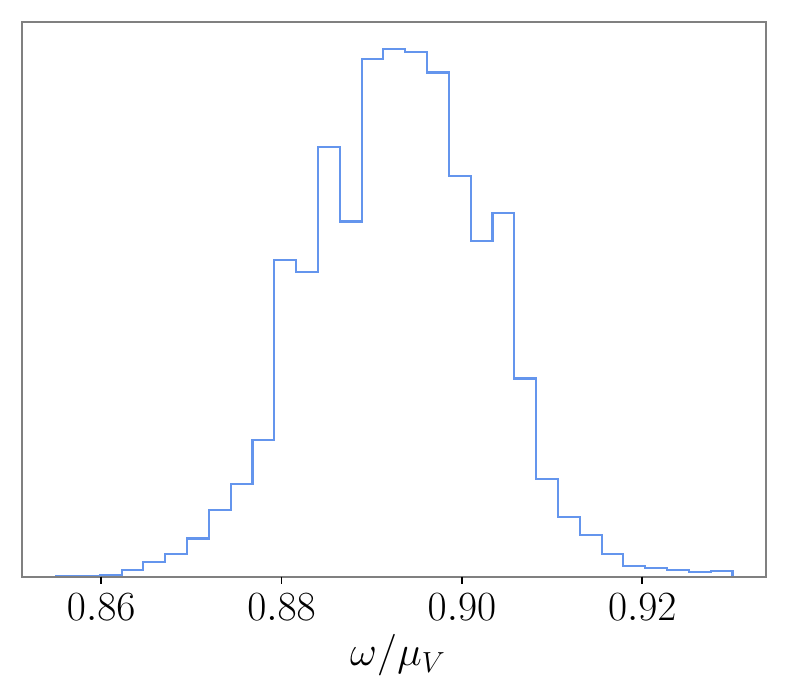}
\includegraphics[width=0.22\textwidth]{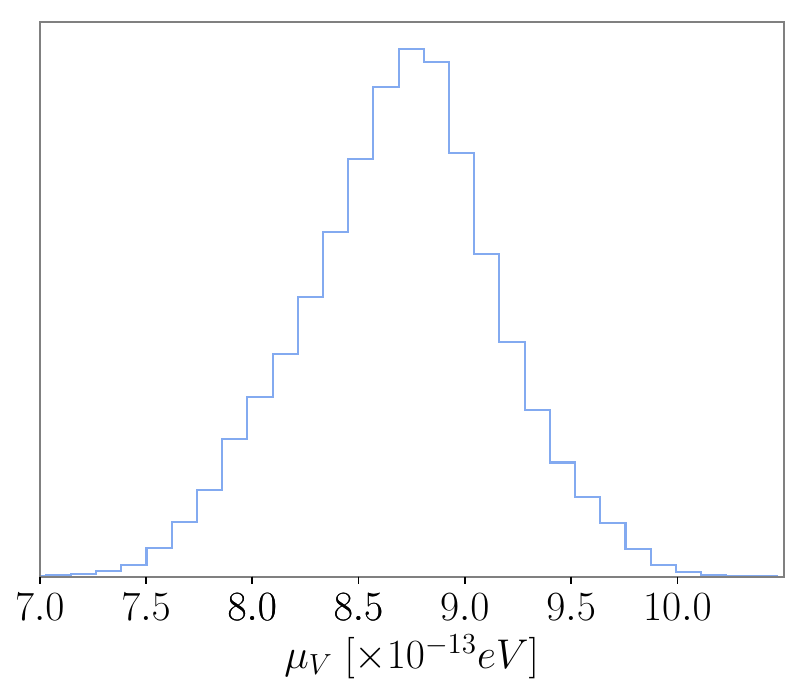}
\caption{\textbf{Posterior distribution for the values of the bosonic field associated with GW190521.} The left panel shows the oscillation frequency of the bosonic field $\omega/\mu_V$. The right panel shows the mass of the ultra-light boson $\mu_V$. We assume a merger of two equal-mass and equal-spin Proca stars.}
\label{fig:final}
\end{center}
\end{figure}

We investigate the physical properties of the hypothetical bosonic field encoded in GW190521. Fig.~\ref{fig:final} shows our posterior distributions for the oscillation frequency (normalized to the boson mass) and for the boson mass $\mu_V$ itself. We constrain the former to be $\omega/\mu_V = 0.893^{+0.015}_{-0.015}$. 
%

To obtain the boson mass $\mu_V$ one must recall that each PS model is characterized by a dimensionless mass $\mathcal{M}_{\text{PS}}=M_{\text{PS}}\, \mu_{V}/M_{\text{Pl}}^2$, with $M_{\text{Pl}}$ the Planck mass. 
Identifying $M_{\text{PS}}$ with half the mass of the final BH in GW190521 we obtain
%
\begin{equation}
    \mu_V=1.34\times10^{-10}\,\biggl(\frac{\mathcal{M}_{\text{PS}}}{M_{\rm BH}^{\rm final}/2}\biggl)\,\text{eV},
\end{equation}
%
%
where $M_{\rm BH}^{\rm final}$ should be expressed in solar masses.  
This yields $\mu_V=8.72^{+0.73}_{-0.82}\times10^{-13}$\,\text{eV}.\\

Finally, we estimate the maximum possible mass for a PS described by such ultra-light boson using

\begin{equation}
    \biggl(\frac{M_\text{max}}{M_{\odot}}\biggl)=1.125\,\biggl(\frac{1.34\times10^{-10}\,\text{eV}}{\mu_V}\biggl).
\end{equation}
This yields $M_\text{max}=173^{+19}_{-14}\, M_{\odot}$. Binaries with lower total masses than this $M_\text{max}$ would produce a remnant that would not collapse to a BH; therefore, they would not emit a ringdown signal mimicking that of a BBH. 
We therefore discard PSs characterised by the above $\mu_V$ as sources of any of the previous Advanced LIGO - Virgo BBH observations, as the largest (redshifted) total mass among these, corresponding  to GW170729, is only around $120\,M_\odot$ \cite{LIGOScientific:2018mvr,Chatziioannou:2019dsz}.


%

While our PHOC analysis is limited to equal-masses and spins, we performed a preliminary exploration of unequal-mass cases. To do this, we fix the primary oscillation frequency to $\omega_1/\mu_V = 0.895$, varying $\omega_2/\mu_V$ along an uniform grid. Table \ref{tab:parameters1} reports our parameter estimates, fully consistent with those for the equal-mass case. 
 We obtain, however, a slightly larger evidence of $\log{\cal{B}}^{\rm PHOC}_{\rm BBH} = 1.9$ that we attribute to the larger distance estimate $d_L = 700^{+292}_{-279}$ Mpc. This indicates that a more in-depth exploration of the full parameter space may be of interest, albeit not impacting significantly on our main findings.

\begin{table}[t!]
\centering
\begin{center}
\begin{tabular}{lcc}
Parameter  & $q=1$ model & $q\neq 1$ model \\ \hline
\\[0pt]
Primary mass & $115^{+7}_{-8}\; M_\odot$ & $115^{+7}_{-8}\, M_\odot$
\\[6pt]
Secondary mass & $115^{+7}_{-8}\; M_\odot$ & $111^{+7}_{-15}\,M_\odot$
\\[6pt]
Total / Final mass & $231^{+13}_{-17}\; M_\odot$ & $228^{+17}_{-15} \,M_\odot$
\\[6pt]
Final spin & $0.75^{+0.08}_{-0.04}$ & $0.75^{+0.08}_{-0.04}$
\\[6pt]
Inclination $\pi/2-|\iota-\pi/2|$ & $0.83^{+0.23}_{-0.47}$ rad &  $0.58^{+0.40}_{-0.39}$ rad 
\\[3pt]
Azimuth  & $0.65^{+0.86}_{-0.54}$ rad &  $0.78^{+1.23}_{-1.20}$ rad
\\[3pt]
Luminosity distance & $571^{+348}_{-181}$ Mpc & $700^{+292}_{-279}$ Mpc 
\\[3pt]
Redshift  & $0.12^{+0.05}_{{-0.04}}$ & $0.14^{+0.06}_{{-0.05}}$
\\[6pt]
Total / Final redshifted mass  & $258^{+9}_{-9}\; M_\odot$ & $261^{+10}_{-11}\; M_\odot$
\\[6pt]
Bosonic field frequency $\omega/\mu_V$  & $0.893^{+0.015}_{-0.015}$ &   $(*) 0.905^{+0.012}_{-0.042}$
\\[3pt]
Boson mass $\mu_V$ [$\times 10^{-13}$] & $8.72^{+0.73}_{-0.82} \;$ eV & $8.59^{+0.58}_{-0.57} \;$ eV 
\\[3pt]
Maximal boson star mass  & $173^{+19}_{-14}$ $M_\odot$ &  $175^{+13}_{-11}$ $M_\odot$
\\[6pt] 
\end{tabular}
\caption{\textbf{Parameters of GW190521 assuming a head-on merger of Proca stars.} \textcolor{black}{In the the first column we assume equal masses and spins while the second corresponds to our exploratory model for unequal masses. There, the asterisk $(*)$ denotes that we estimate the oscillation frequency of the secondary bosonic field $\omega_2/\mu_V$, while that for the primary star is fixed to $\omega_1/\mu_V=0.895$}. We report median values and symmetric $90\%$ credible intervals.} 
\label{tab:parameters1}
\end{center}
\end{table}

 {\it \textbf{Discussion.}} We have compared GW190521 to numerical simulations of BH head-on mergers and horizonless bosonic stars known as PSs. While we discard the first scenario, we have shown that GW190521 is consistent with an equal-mass head-on merger of PSs, inferring an ultralight boson mass $\mu_V \simeq 8.72\times10^{-13}$ eV. 
%


Current constraints on the boson mass are obtained from the lack of GW emission associated with the superradiance instability and from observations of the spin of astrophysical BHs \cite{baryakhtar2017black,cardoso2018constraining,palomba2019direct}.
These, however, apply to \textit{real} bosonic fields. For complex bosonic fields, the corresponding cloud around the BH does not decay by GW emission, but a stationary and axisymmetric Kerr BH with bosonic hair forms \cite{herdeiro2014kerr,herdeiro2016kerr,east2017superradiant}. These configurations are, themselves, unstable against superradiance \cite{ganchev2018scalar}, but the non-linear development of the instability is too poorly known to establish meaningful constraints on the complex bosons - see, however \cite{degollado2018effective}.

\textcolor{black}{Our study is limited to head-on mergers of bosonic stars due to the current lack of methods to simulate less eccentric configurations. Remarkably, however, these suffice to fit GW190521 as closely as state-of-the art BBH models, being slightly favoured from a Bayesian point of view. While this restriction leads to narrow parameter distributions, the future development of more complex configurations like quasi-circular mergers shall reveal, for instance, a larger range of boson masses consistent with GW190521. This could potentially reduce the corresponding bound on the maximum mass of a stable boson star, $M_{\rm max}$, and make some of the previous LIGO-Virgo events candidates for mergers of Proca stars with a compatible boson-mass $\mu_V$. To numerically simulate such configurations, however, constraint-satisfying initial data are needed to obtain accurate waveforms, which are currently unavailable. We believe that our results will strongly motivate efforts to build such initial data. 
}


The existence of an ultra-light bosonic field would have profound implications. It could account for, at least, part of dark matter, as it would give rise to a remarkable energy extraction mechanism from astrophysical spinning BHs,  eventually forming new sorts of  ``hairy" BHs. In addition, such field could serve as a guide towards beyond-the-standard-model physics, possibly pointing to the stringy  axiverse.

While GW190521 does not allow to clearly distinguish between the BBH and PS scenarios, future GW observations in the IMBH range shall allow to better resolve the nature of the source, helping confirm or reject the existence of the ultra-light vector boson discussed here.

\section*{Acknowledgements \\}

The authors thank Fabrizio Di Giovanni, Tjonnie G.F. Li and Carl-Johan Haster for useful discussions and Archana Pai for comments on the manuscript. The analysed data and the corresponding power spectral densities are publicly available at the online Gravitational Wave Open Science Center \cite{GWOSC}. LVC parameter estimation results quoted throughout the paper and the corresponding histograms and contours in Fig. 2 have made use of the publicly available sample release in \texttt{https://dcc.ligo.org/P2000158-v4}. JCB is supported by the Australian Research Council Discovery Project DP180103155 and by the Direct Grant from the CUHK Research Committee with Project ID: 4053406. The project that gave rise to these results also received the support of a fellowship from ”la Caixa” Foundation (ID
100010434) and from the European Union’s Horizon
2020 research and innovation programme under the
Marie Skłodowska-Curie grant agreement No 847648.
The fellowship code is LCF/BQ/PI20/11760016. JAF is supported by the Spanish Agencia Estatal de Investigaci\'on  (PGC2018-095984-B-I00) and by the  Generalitat  Valenciana  (PROMETEO/2019/071). This work is supported by the Center for Research and Development in Mathematics and Applications (CIDMA) 
through the Portuguese Foundation for Science and Technology (FCT - Funda\c {c}\~ao para a Ci\^encia e a Tecnologia), references UIDB/04106/2020, UIDP/04106/2020, UID/FIS/00099/2020 (CENTRA), and by national funds (OE), through FCT, I.P., in the scope of the framework contract foreseen in the numbers 4, 5 and 6 of the article 23, of the Decree-Law 57/2016, of August 29, changed by Law 57/2017, of July 19.
We also acknowledge support  from  the  projects  PTDC/FIS-OUT/28407/2017,   CERN/FIS-PAR/0027/2019 and PTDC/FIS-AST/3041/2020.   
This work has further been supported by the European Union’s Horizon 2020 research and innovation (RISE) programme H2020-MSCA-RISE-2017 Grant No. FunFiCO-777740.  The authors would like to acknowledge networking support by the COST Action CA16104.
The authors acknowledge computational resources provided by the LIGO Laboratory and supported by
National Science Foundation Grants PHY-0757058 and PHY0823459; and the support of the NSF CIT cluster for the provision of computational resources for our parameter inference runs.  This manuscript has LIGO DCC number P-2000353. 

\section*{Supplementary Material}

\section{Source orientation, frame-dragging and the ($l$, $m$)=(2,0) mode}
\begin{figure}[t!]
\begin{center}
\includegraphics[width=0.4\textwidth]{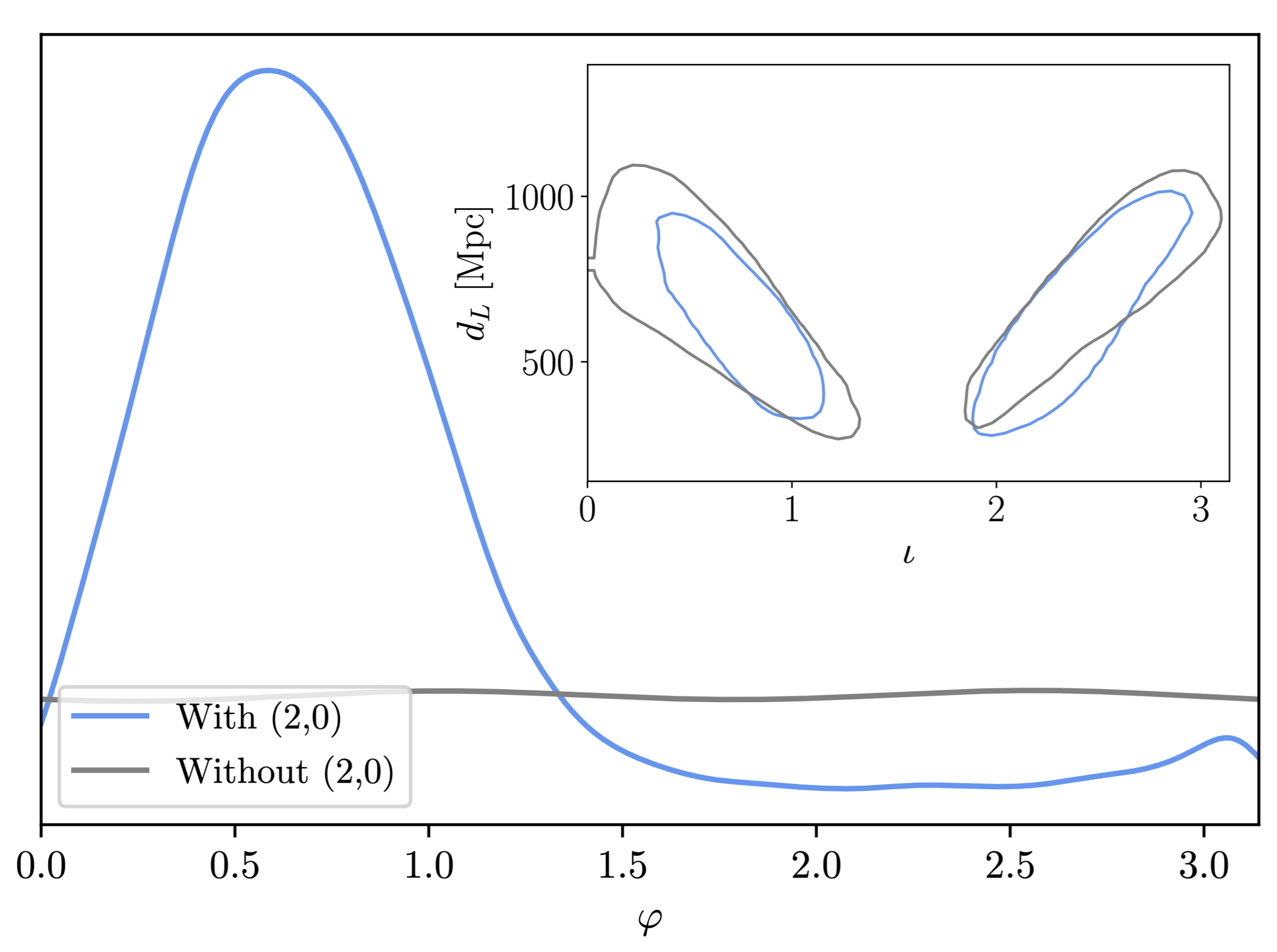}
\includegraphics[width=0.44\textwidth]{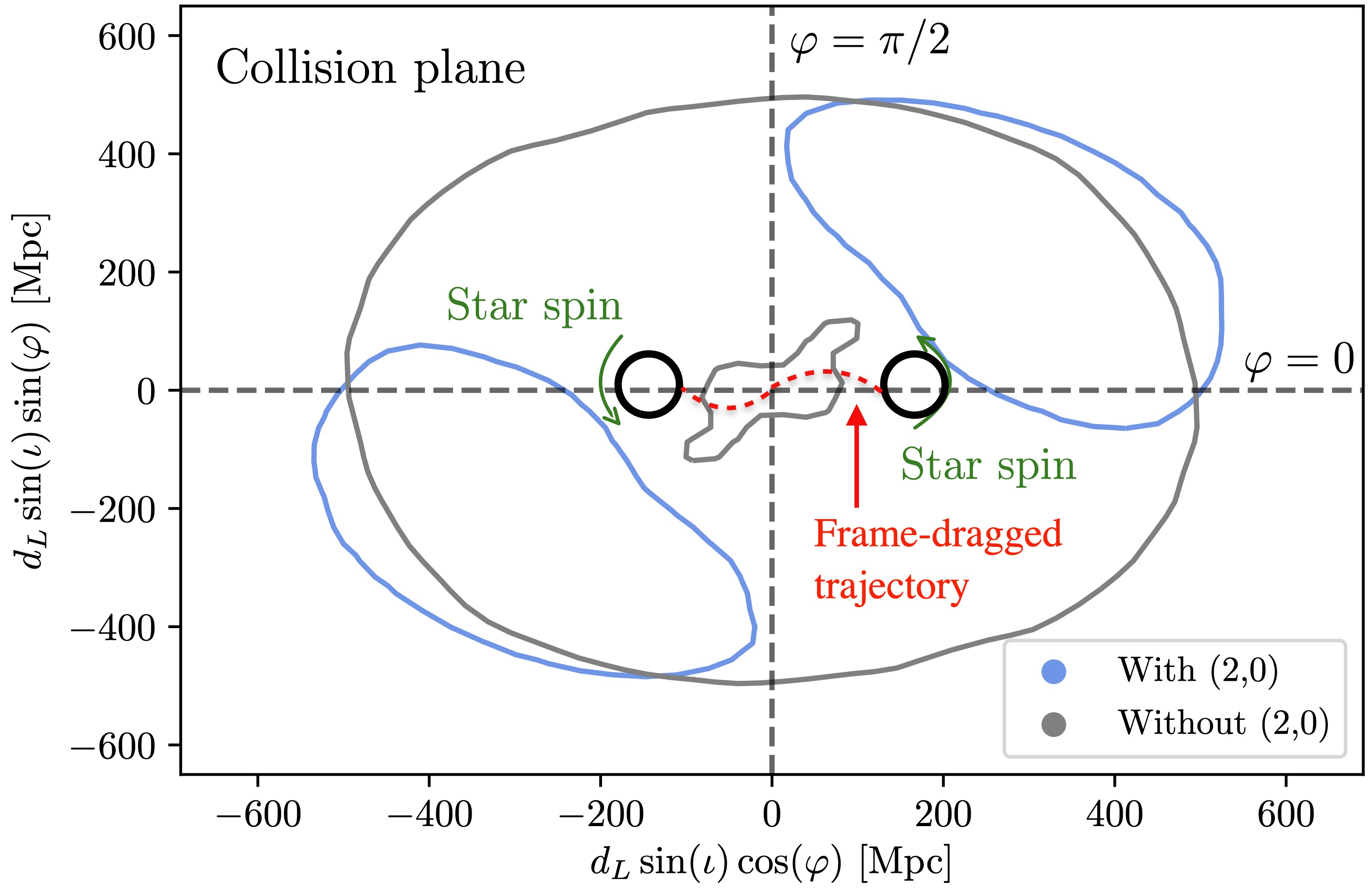}
\caption{\textbf{Distance and orientation of GW190521 under the assumption of an equal-mass, equal-spin PS merger.} \textbf{Left:} The main panel shows the posterior distribution for the azimuthal angle $\varphi$, with $\varphi=0,\pi$ denoting the line joining the two PSs at the start of the simulation and increasing in the direction of the spin of the individual PSs and final BH. The inset shows the two dimensional $90\%$ credible regions for the inclination angle $\iota$ and the luminosity distance $d_L$. The angle $\iota=0$ denotes the direction of the final spin. We show results obtained with templates including and omitting the $(2,0)$ mode. \textbf{Right:} 2-dimensional $90\%$ credible regions for the projection of the location $(d_L,\iota,\varphi)$ of the observer around GW190521 onto the collision plane. The azimuthal angle $\varphi$ grows in the direction of the spin of the stars, while the inclination angle $\iota$ is $0(\pi)$ at the center of the plot and evolves toward $\pi/2$ for increasing magnitude of the x and y axis. When including the $(2,0)$ mode in the analysis, we constrain the azimuth to the first and third quadrants of the collision plane. The trajectories of the two stars curve towards increasing values of $\varphi$ due to frame dragging induced by the spins. The size and separation of the stars has been greatly exaggerated to facilitate the interpretation of the plot.}
\label{fig:angles}
\end{center}
\end{figure}

Here we provide further details on our estimation of the source orientation, denoted by the polar and azimuthal angles $(\iota,\varphi)$ and the impact of the $(2,0)$ mode on this measurement.\\

The inset in the top left of Fig.~\ref{fig:angles} shows our 2-dimensional $90\%$ credible region for the distance $d_L$ and inclination angle $\iota$ of GW190521, assuming a PHOC scenario. The latter is defined as the angle formed by the final spin and the line-of-sight. Contrary to quasi-circular mergers, the GW emission from head-on mergers is not dominated by the quadrupole $(2,2)$ mode \cite{maggiore2008gravitational}, but have a similarly strong $(2,0)$ mode \cite{anninos1995head,palenzuela2007head}. This provides a richer morphology to the signals \cite{Graff:2015bba,CalderonBustillo:2018zuq,CalderonBustillo:2019wwe} that breaks degeneracies between parameters, e.g., that between the distance and the inclination angle \cite{Graff:2015bba,Chatziioannou:2019dsz,LIGOScientific:2020stg} and that between the polarisation angle and the azimuth \cite{CalderonBustillo:2018zuq}. The inclusion of this mode in our templates helps to better constrain not only the distance and orientation of the binary but also allows to estimate, for the first time, the azimuthal angle $\varphi$ describing the location of the observer around the source.

We find that the inclusion of the $(2,0)$ mode disfavours face-on(off) orientations given by $\iota=0 (\pi)$, for which this mode is suppressed, hence suggesting its presence in the signal. By repeating our analysis excluding the $(2,0)$ mode from our templates, we obtain a $\log{\cal B}\sim 0.6$, mildly favouring the presence of the $(2,0)$ mode. The inclination of the source, together with the asymmetry in the GW emission produced by this mode, allows to measure the azimuthal angle of the observer, understood as that formed by the collision axis and the projection of the line-of sight onto the plane normal to the final spin. We constrain this to $0.06 < \varphi < 1.40$ radians (see main left panel of Fig.~\ref{fig:angles}), measured in the direction of any of the two PSs spins. To facilitate an interpretation of this measurement, the right panel Fig. \ref{fig:angles} shows the $90\%$ credible intervals for the projection of the location of the observer around the source $(d_L,\iota,\varphi)$ (or conversely, distance and source orientation) onto the collision plane. We restrict $\varphi$ to the first and third quadrants of this plane. This can be physically interpreted as the trajectory of closest PS to Earth being curved towards us due to the frame-dragging induced by the spins, while the furthest one curves away. To the best of our knowledge, this is the first time such measurement is performed.\\

\section{Results for alternative distance prior}

Standard parameter estimation assumes that gravitational-wave sources distribute uniformly in co-moving volume. While this is a sensible assumption, it does intrinsically favour loud sources like quasi-circular BH mergers, over weaker head-on BH mergers. In this section we investigate the impact of this prior on the results presented in the main text by repeating our analysis imposing a prior uniform in distance, and report the Log Bayes factors obtained for each of our models
~(see Table~\ref{tab:logb}). As it is evident, a prior uniform in distance removes the intrinsic bias for loud sources, giving significantly larger evidences for the PHOC models than in the main text. Table ~\ref{tab:parameters} shows the corresponding parameter estimates, fully consistent with those obtained using the standard distance prior, albeit slighlty more noticeable (and expected) changes in the distance and redshift (to lower values), and the source-frame mass (to larger values). In particular, we obtain fully consistent results for the frequency and the particle mass characterising the bosonic field, as well as for the maximum PS mass.

\begin{table}[t!]
\centering
\begin{center}
\begin{tabular}{lcc}
\rule{0pt}{3ex}
Waveform Model & $\log{\cal B}$ &  $\log{\cal L}_{Max}$ \\
\hline
\rule{0pt}{3ex}%
Quasi-circular Binary Black Hole & 80.1 & 105.2
\\
\rule{0pt}{3ex}%
Head-on Equal-mass Proca Stars &  83.5  & 106.7\\
\rule{0pt}{3ex}%
Head-on Unequal-mass Proca Stars &   84.3 & 106.5\\
\rule{0pt}{3ex}%
Head-on Binary Black Hole &  78.0 &  103.2 \\ 
\end{tabular}

\caption{\textbf{Bayesian evidence for our source models using a uniform-in-distance prior:} The first four rows show the natural Log Bayes Factor obtained for our different waveform models and corresponding maximum values of the Log Likelihood. 
}
\label{tab:logb}
\end{center}
\end{table}

\begin{table}[t!]
\centering
\begin{tabular}{lc}
Parameter  & \\ \hline
\\[0pt]
Total / Final mass &  $234^{+12}_{-16}\; M_\odot$
\\[6pt]
Final spin &  $0.75^{+0.08}_{-0.04}$
\\[6pt]
Inclination & $0.88^{+0.24}_{-0.35}$ rad 
\\[3pt]
Azimuth  &  $0.60^{+0.66}_{-0.48}$ rad
\\[3pt]
Luminosity distance & $520^{+274}_{-181}$ Mpc
\\[3pt]
Redshift   & $0.10^{+0.03}_{{-0.05}}$
\\[6pt]
Total / Final redshifted mass  & $259^{+8}_{-10}\; M_\odot$
\\[6pt]
Frequency of bosonic field &  $0.892^{+0.016}_{-0.020}$
\\[3pt]
Boson mass & $8.65^{+0.77}_{-0.84} \;$ eV 
\\[3pt]
Maximal boson star mass  & $174^{+19}_{-12}$ $M_\odot$
\\[6pt]
\end{tabular}
\caption{\textbf{Parameters of GW190521 assuming an equal-mass head-on merger of PSs and an uniform distance prior.} The result are consistent with those reported in Table II in the main text, despite expected differences in the luminosity distance, redshift and source-frame mass.}
\label{tab:parameters}
\end{table}

\section{Numerical Simulations}
Bosonic stars, and in particular Proca stars, are fundamentally different from BHs and neutron stars. For the former the angular momentum $J$ can vary continuously for a given mass $M$. In contrast, in the bosonic case, the value $\omega/\mu_V$ determines the mass (as function of $\mu_V$) and the compactness of the model, but also the angular momentum. Therefore, a given value of $\omega/\mu_V$ only allows one $(M, J)$ pair. In addition, the angular momentum is quantized and determined by an integer $m$, the azimuthal angular momentum number. This property reduces the space of parameters of bosonic stars. While $m=0$ corresponds to the non-spinning solutions, models with $m\geq2$ are unstable against non-axisymmetric perturbations \cite{SanchisGual2019}. Therefore, we restrict our study to PSs with $m=1$.

\begin{figure}[h!]
\begin{center}
\includegraphics[width=0.19\linewidth]{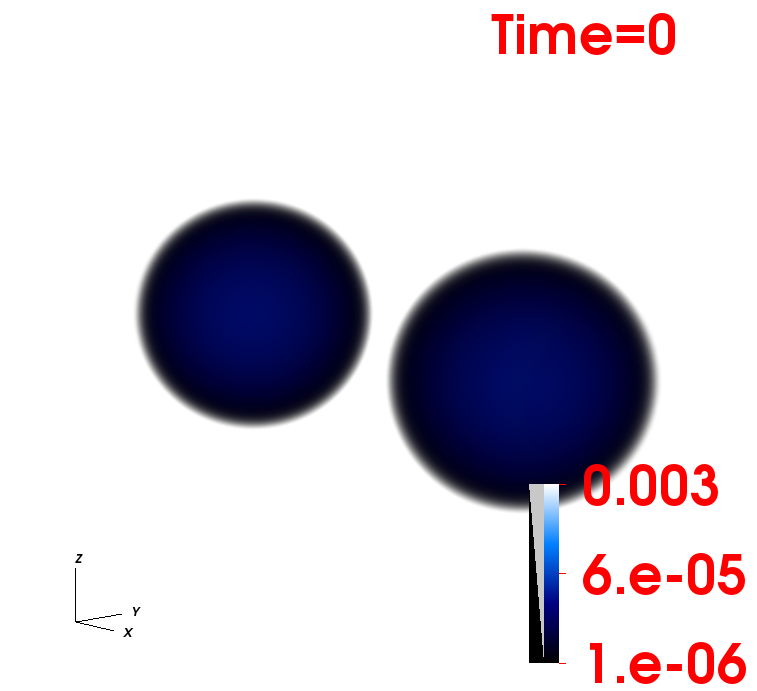}\hspace{-0.007\linewidth}
\includegraphics[width=0.19\linewidth]{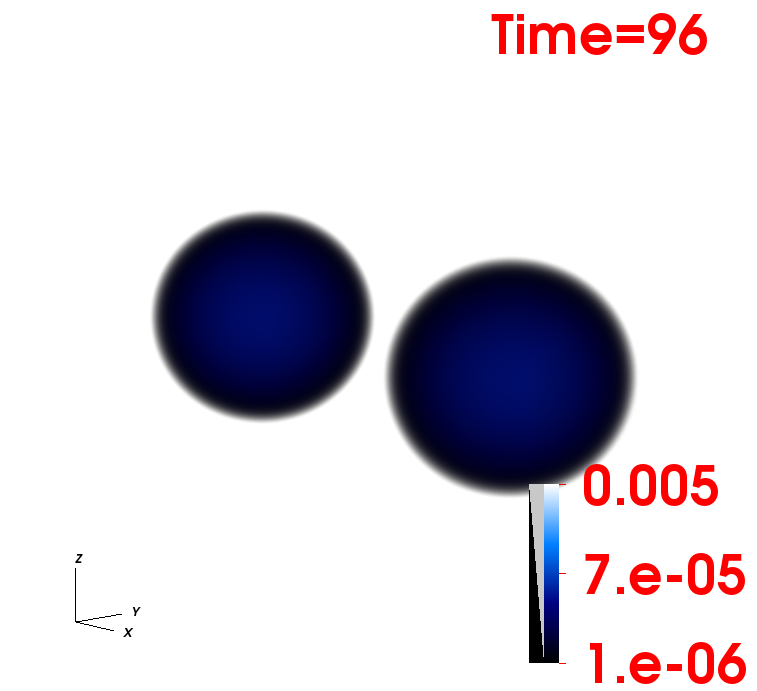}\hspace{-0.007\linewidth}
\includegraphics[width=0.19\linewidth]{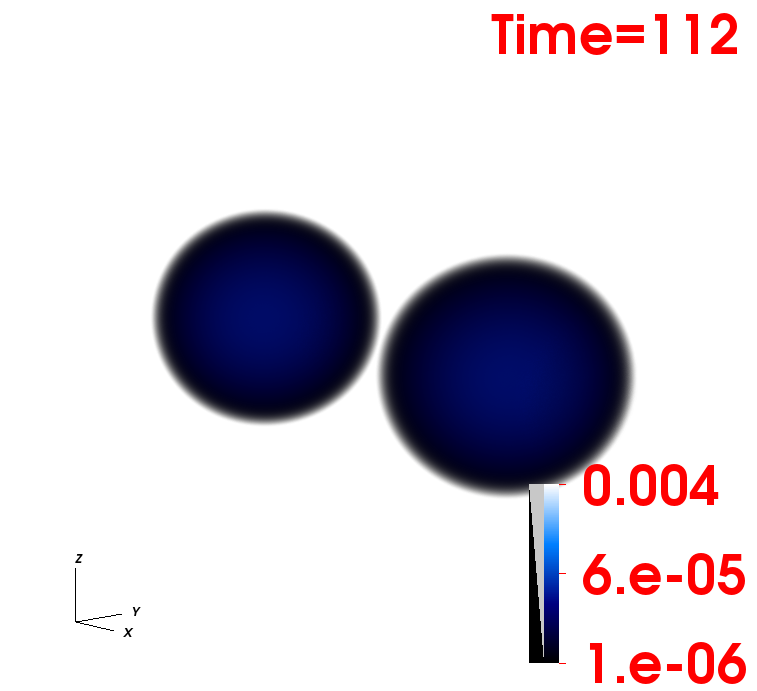}\hspace{-0.007\linewidth}
\includegraphics[width=0.19\linewidth]{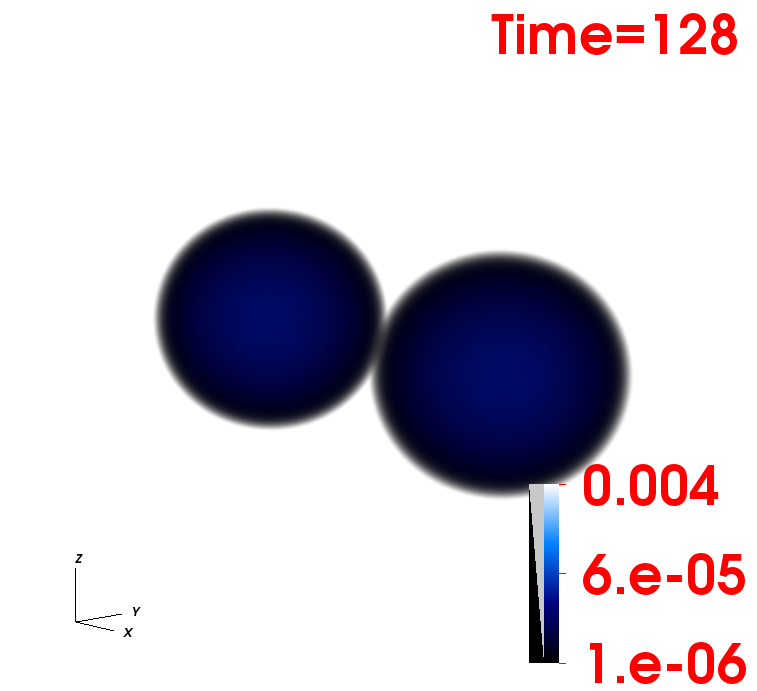}\hspace{-0.007\linewidth}
\includegraphics[width=0.19\linewidth]{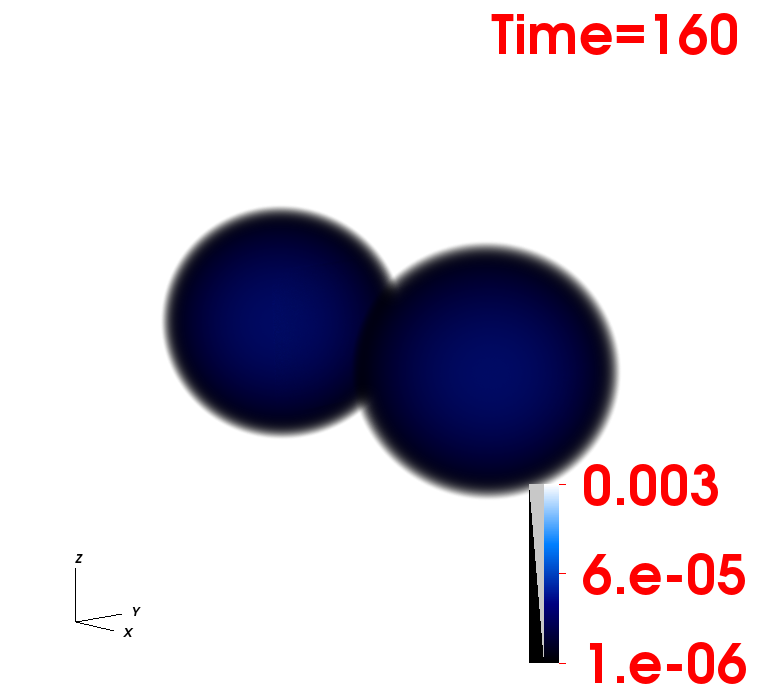}\\
\includegraphics[width=0.19\linewidth]{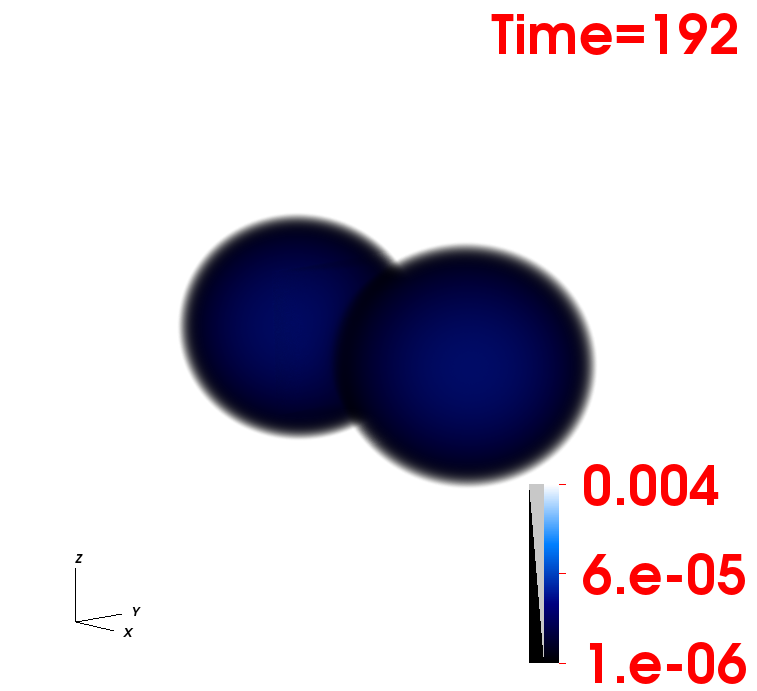}\hspace{-0.007\linewidth}
\includegraphics[width=0.19\linewidth]{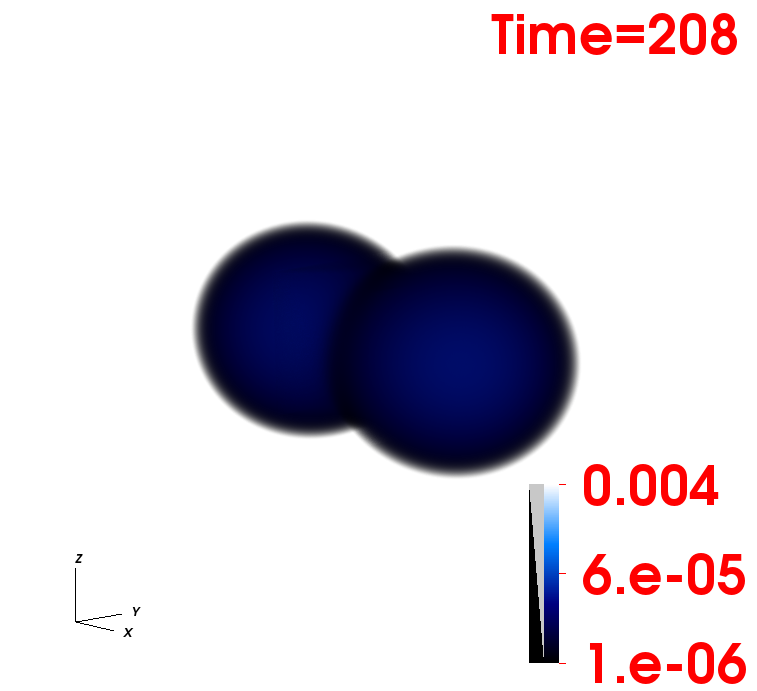}\hspace{-0.007\linewidth}
\includegraphics[width=0.19\linewidth]{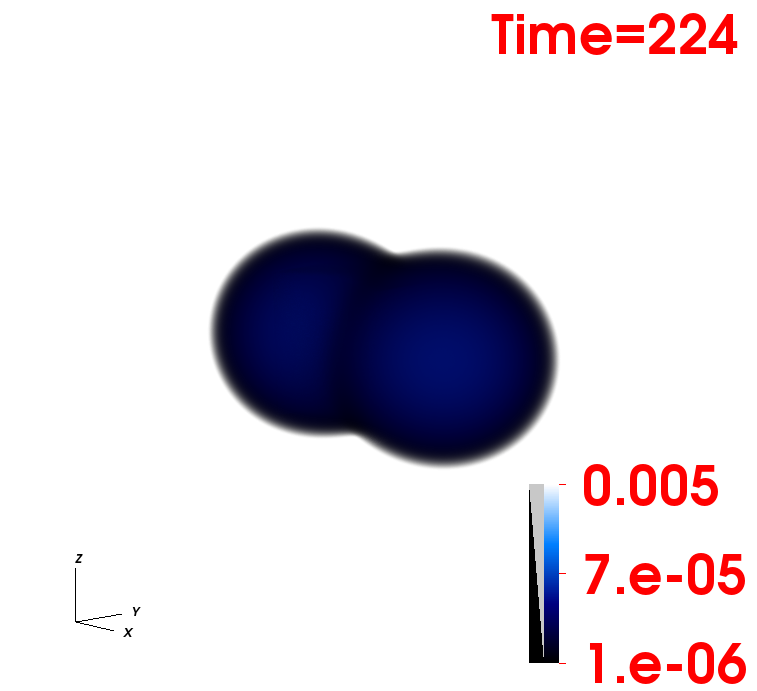}\hspace{-0.007\linewidth}
\includegraphics[width=0.19\linewidth]{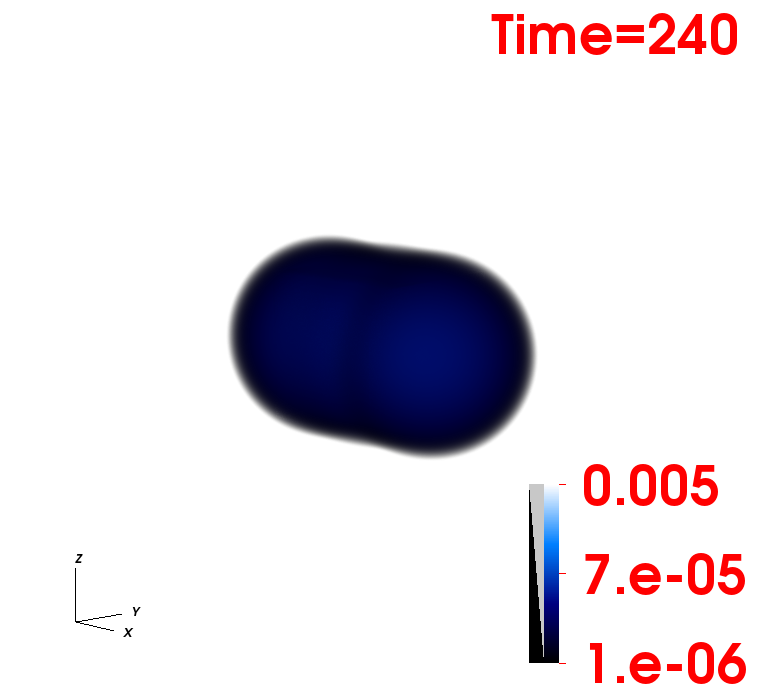}\hspace{-0.007\linewidth}
\includegraphics[width=0.19\linewidth]{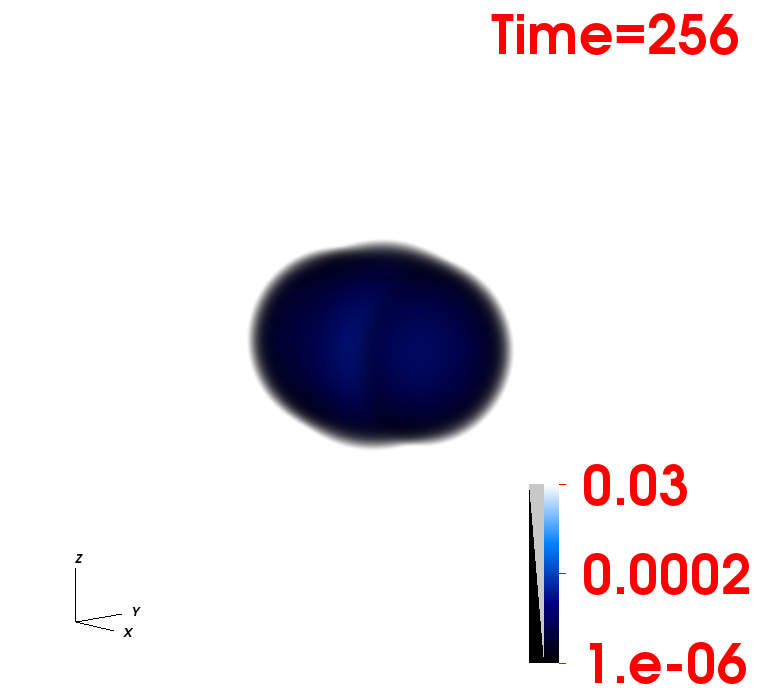}\\
\includegraphics[width=0.19\linewidth]{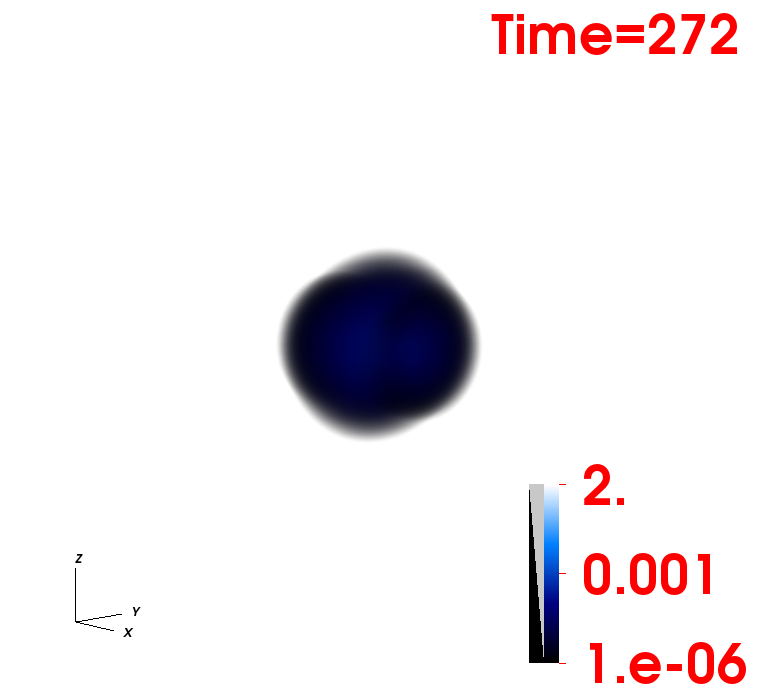}\hspace{-0.007\linewidth}
\includegraphics[width=0.19\linewidth]{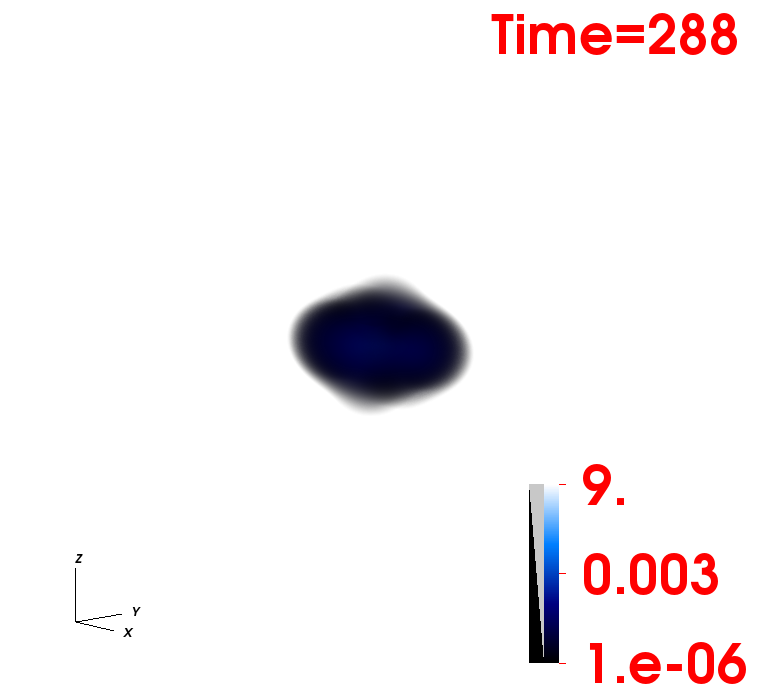}\hspace{-0.007\linewidth}
\includegraphics[width=0.19\linewidth]{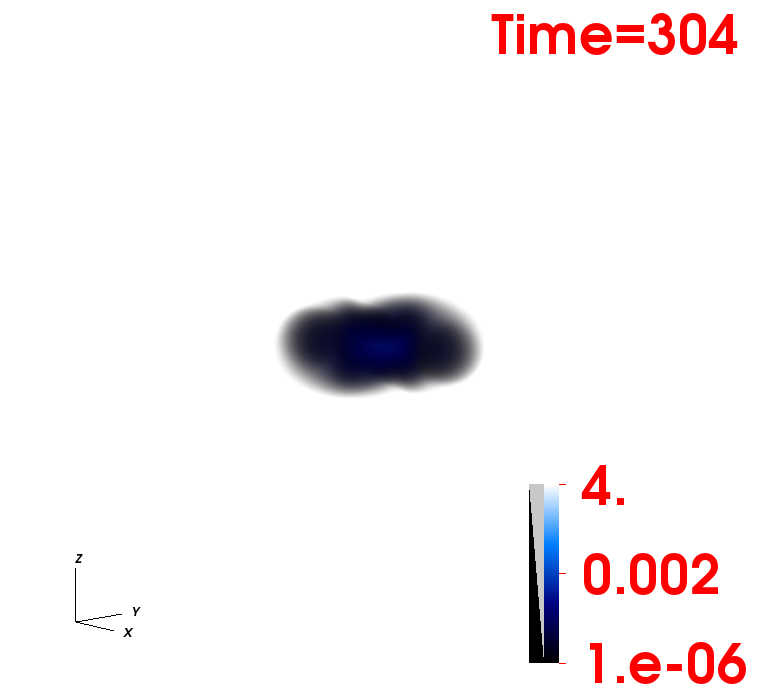}\hspace{-0.007\linewidth}
\includegraphics[width=0.19\linewidth]{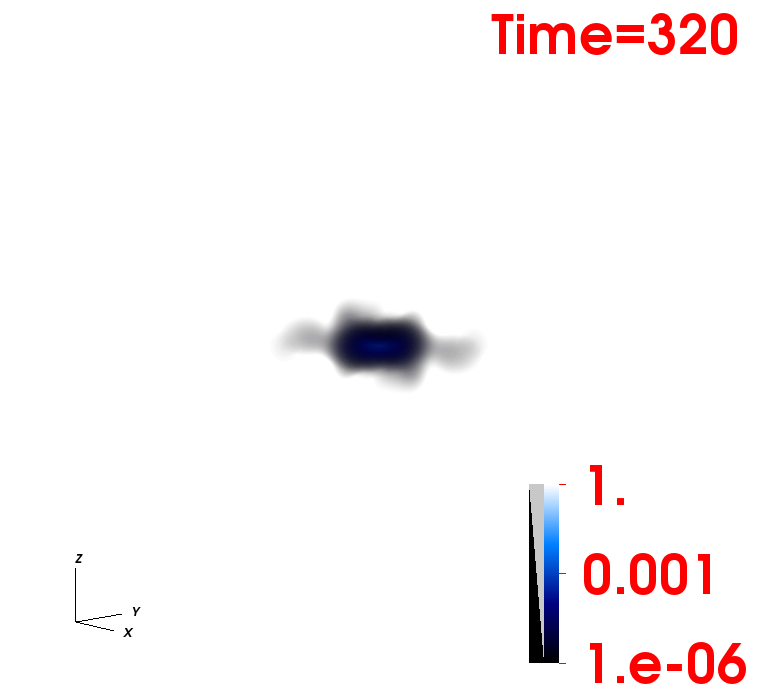}\hspace{-0.007\linewidth}
\includegraphics[width=0.19\linewidth]{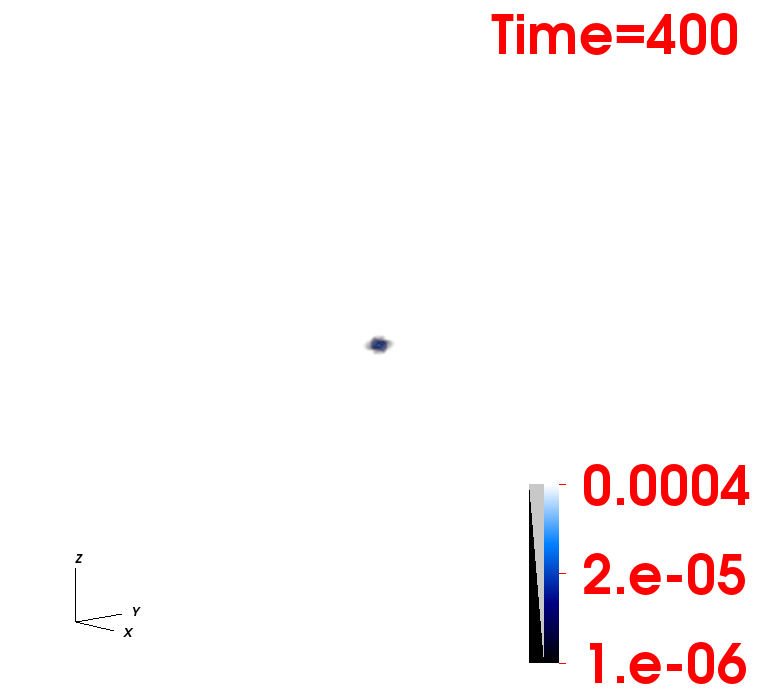}
\caption{\textbf{Snapshots of the time evolution of the energy density during the head-on collision of two PSs with $\omega/\mu_V=0.8925$.} Time is given in code units.}
\label{fig:proca}
\end{center}
\end{figure}

To perform the numerical evolutions we have used the \etk infrastructure \cite{EinsteinToolkit,Loffler:2011ay, Zilhao:2013hia} with the \carpet package \cite{Schnetter:2003rb,Cactus} for mesh-refinement capabilities. The initial data for the BH head-on collision are calculated using the \textsc{TwoPunctures} thorn \cite{ansorg2004single}. The Proca equations are solved using a modification of the \textsc{Proca} thorn \cite{Canuda_2020_3565475, Zilhao:2015tya} to include a complex field. We have performed numerical simulations of head-on collisions of equal mass PSs. The initial data consists in the superposition of two solutions separated by $D = 40/\mu_V$, in geometric units, to reduce the initial constraint violations. In total we have evolved 77 models with different frequency $\omega$, total mass, angular momentum and compactness. In Fig.~\ref{fig:proca}, we show the time evolution of the energy density for a PS with $\omega/\mu_V=0.8925$. 

Numerical simulations extract the gravitational-wave emission in  terms  of  the  Weyl  curvature  component, $\psi_4$. Therefore, it is necessary to integrate twice in time to obtain the strain $h$. This process is not trivial and can produce non-linear drifts in the resulting strain \cite{Reisswig:2015}. To avoid these issues, we integrate the $\psi_4$ component in frequency domain, introducing a small regularization term to avoid the singularity at $f=0$ Hz. Then we apply a high-pass filter to reduce the energy contained in frequencies lower than a chosen cutoff.

\section{Parameter Estimation} 

In this section we provide details regarding our analysis set-up and our parameter estimation procedure. In particular, we explain in detail how continuous parameter distributions are obtained from a set of discrete simulations for PSs and BH head-on mergers, as well as the corresponding evidences and Bayes' Factors. We also discuss the comparison of the evidence for these PS models to that of BBHs.

\subsection{Data}

We perform \textcolor{black}{full Bayesian} parameter estimation on GW190521 using the software \texttt{bilby} \cite{Ashton:2018jfp} together with the \texttt{cpnest} nested sampling algorithm \cite{CPNest}. We analyse four seconds of publicly available data around the trigger time of GW190521  \cite{GWOSC}, sampled at 1024 Hz, using the corresponding power spectral density computed by BayesWave \cite{Cornish:2014kda,Cornish2015}.\\

\subsection{Summary of Bayesian Inference}

The posterior probability for a set of source parameters $\theta$, given a stretch of data $d$, is given by \begin{equation}
p(\theta| d,  {\cal{M}}) = \frac{\pi(\theta) {{\cal{L}}(d|\theta,{\cal{M}})} }{{\cal{Z}}(d|{\cal{M}})},
\label{eq:bayes}
\end{equation}
where ${\cal{ L}}$ denotes the standard frequency-domain likelihood commonly used for gravitational-wave transients \cite{Finn1992,Romano2017}
\begin{equation}
 \log{\cal{L}}(d|\theta,{\cal{M}}) =  - \frac{1}{2}(d-h_{{\cal{M}}}(\theta)|d-h_{{\cal{M}}}(\theta)).
\end{equation}
Here, $h(\theta)$ denotes a waveform template for parameters $\theta$, according to the waveform model ${\cal{M}}$. In our case we consider three models, respectively representing quasi-circular binary black hole mergers (BBH), head-on black-hole mergers (HOC) and head-on Proca Star mergers (PHOC). As usual, the operation $(a|b)$ denotes the inner product \cite{Cutler1994}
\begin{equation}
    (a|b)= 4 \Re \int_{f_{\min}}^{f_{\max}} \frac{\tilde{a}(f)\tilde{b}(f)}{S_n(f)} df,
\end{equation}
where $S_n(f)$ denotes the one sided power spectral density of the detector noise, and $f_{\min}$ and $f_{\max}$ are respectively the low and high frequency cutoffs of the detector data. The factor $\pi(\theta)$ denotes the prior probability for the parameters $\theta$ and the factor ${\cal{Z}}(d|{\cal{M}})$ is known as the evidence for the model ${\cal{M}}$. This is given by the integral of the numerator of Eq.\ref{eq:bayes} across all the parameter space covered by the model
\begin{equation}
{\cal{Z}}(d|{\cal{M}}) := {{\cal{Z}}}_{{\cal{M}}} = \int \pi(\theta) {{\cal{L}}(d|\theta,{\cal{M}})} d\theta.\\
\label{eq:z}
\end{equation}

Given two models $A$ and $B$, the degree of preference for model $A$ over model $B$ is given by the Bayes' Factor
\begin{equation}
{\cal{B}}^{A}_{B} = \frac{{\cal{Z}}_A}{{\cal{Z}}_B}.
\end{equation}
Throughout the main text, we refer to $\log({\cal{Z}}_A)$ and $\log({\cal{Z}}_B)$ as the ``Log Bayes Factor'' for each of the models (with respect to the noise, i.e., no-signal hypothesis) and to $\log({\cal{Z}}_A) - \log({\cal{Z}}_B)$ as the relative Log Bayes Factor $\log{\cal{B}}^{A}_B$.
It is common to say that the model $A$ is strongly preferred wrt. $B$ when the log ${\cal{B}}^{A}_{B} > 5$. This is, when model $A$ is $\sim 150$ times more probable than model $B$.\\

\subsection{Prior choices}

Our three models cover different parameter spaces. As usual, the BBH model covers a continuous 15-dimensional parameter space $\Theta$ formed by the total mass $M$, the mass ratio $q$, the six individual spin components $(\vec a_1, \vec a_2)$, the two orientation angles $(\iota,\varphi)$, the two sky-localisation angles, the luminosity distance $d_L$, the polarisation angle and the time of arrival. However, our HOC and PHOC models cover only a discrete set of spins and mass ratios, sharing all the other parameters with the BBH model, making the analysis less trivial.\\

For the case of the HOC and PHOC models we cannot sample over the two individual masses of the binary (as it is common practice) as our simulations only cover a discrete range of mass ratios. Since these simulations scale trivially with the total mass, it is natural to place an uniform prior on it. We choose an uniform prior in $[100,500]M_\odot$ for all of our models. In addition, we place standard priors on the source orientation, sky-location and polarisation angles. Our PHOC simulations are restricted to mass-ratio $q=1$ and equal-spins, while for the BBH case we place an uniform prior in mass-ratio together with the usual isotropic prior for the two individual spins. For the HOC case, our simulations distribute in a non-uniform way in both mass-ratio and spins, as we produced them in a systematic way trying to maximise the likelihood (see more details in the subsection 6. 
``Evidence for Head-on BBHs'').\footnote{Given the low Bayes' Factors obtained for HOCs, this does not have any impact on the conclusions of our study.} Finally, as we indicate in the main text, we explore two different distance priors. The first one assumes an uniform distribution of sources in co-moving volume. Since this prior will favour large distances, it will prefer loud sources over weak ones, even if both can fit the data equally well. While this is reasonable and also common practice, we try to gauge this away by using also a prior uniform in distance that does not favor loud sources. Finally, we sample the parameter space using the algorithm CPNest \cite{CPNest} and set minimum and maximum frequencies of $11$ and $512$ Hz for our analysis.

\subsubsection{Computing evidences and Bayes Factors for Proca Stars}

Since the BBH model covers a continuous parameter space, it is trivial to compute the integral in Eq.6 across all the space $\bar \Theta$. However, for the case of HOCs and PHOCs we obtain a discrete set of evidences ${\cal{Z}}_{\bar\Theta}$ for each set of mass-ratios and spins, which we shall collectively denote as $\bar \Theta$. In order to find the evidence corresponding to these models, we need to chose a suitable integration element $d \bar \Theta$ to perform the discrete integration over these parameters. While this is intricate for the case of HOC, which we discuss later, for the specific case of PHOCs we can take advantage of the extra parameter $\omega/\mu_V$ that describes the oscillation frequency of the bosonic field.  Since our simulations span an uniform grid in this parameter, we compute the corresponding global evidence as:
\begin{equation}
\begin{aligned}
    \mathcal{Z}_{\text{PHOC}} \approx \sum {\cal{Z}}_{\omega} d\omega = \sum \mathcal{L}(d|\omega, \text{PHOC})\, \pi(\omega)\, d\omega,
\end{aligned}
\end{equation}
with
\begin{equation}
    \mathcal{L}(d|\omega, \text{PHOC}) = \int {\cal{L}}(d|\omega,\hat \theta, \text{PHOC})\, \pi{(\hat \theta)}\, d\hat \theta.
\end{equation}
Here, $\hat \theta$ denotes the extrinsic parameters plus the total mass, so that $\Theta = \bar \Theta \cup \hat \theta$. 

\subsection{The size of the parameter space and the Occam factor}

When comparing the BBH and PHOC models, it is important to note that two main factors determine the value of the corresponding evidences. The first one is how well the model can fit the data. Parameters yielding good fits will yield large values of $\log{\cal{L}}$, and vice versa. In particular, note that ${\cal{Z}}$ is bounded by, e.g.,
\begin{equation}
{\cal{Z}}_{\cal{M}} \leq \int \pi(\theta) {\cal{L}}_{Max} d\theta. 
\end{equation},
with $\log{\cal{L}}_{Max}$ denoting the maximum value of the likelihood across the parameter space. Second, the act of integrating implies that the model may explore regions of the parameter space with poor contributions to the integral. Since $\int\pi(\theta) d\theta = 1$ 
exploring ``useless'' portions of the parameter space leading to poor fits causes to a reduction of ${\cal{Z}}_{\cal{M}}$. This penalty is known as the \textit{Occam factor}. \\ 

Because of our limited computational resources, we only performed enough PS simulations to reconstruct the full posterior distribution for the parameter $\omega/\mu$, shown in Fig.4. As a consequence, we are not exploring a vast parameter space that may provide poor fits to the data, somewhat minimising the Occam Factor and somehow optimising ${\cal{Z}}_{\text{PHOC}}$. Meanwhile, the BBH model covers all the parameter space allowed by the model, leading to an increased Occam Factor and a consequent reduction of ${\cal{Z}}_{\text{BBH}}$. Here we explore some simplifications of the BBH model that shall reduce the Occam penalty, potentially increasing the evidence for the model and reduce the relative evidence in favour of our PHOC model. The results are summarised in Table
~\ref{tab:BBH_occam}, and we describe them in the following.

\begin{enumerate}
    \item \textbf{Aligned spins:} The model NRSur7dq4 includes the effect of orbital precession. This effect is described by the 6 spin components of the two BHs, which greatly increases the explored parameter space wrt., that of our PHOC model. We study the effect of restricting the spins to be (anti-)aligned with the orbital angular momentum, therefore removing the impact of precession and eliminating 4 parameters. Doing so, we find $\log{{\cal{B}}}
   ^{\text{BBH}}_{\text{BBH,AS}} = 3$ \footnote{This corresponds to a $\log_{10}{\cal{B}}
^{\text{BBH}}_\text{{BBH,AS}}\simeq 1.3$. The LVC reported $1.06$ \cite{GW190521D}. Note, however, that while the LVC uses a prior uniform in component masses, we use a prior uniform in total mass and mass ratio.}, accompanied with a much reduced maximum log-likehood of $98.8$. This shows that spin-precession adds a \textit{necessary} complication to the model. Removing this effect increases the evidence for PHOC to $\log{{\cal{B}}}
   ^{\text{PHOC}}_{\text{BBH,AS}} = 3.8$. 
   
   \item \textbf{Mass ratio:} The model NRSur7dq4 covers the mass-ratio range $q \in [1,4]$. However, the LVC results show that mass ratios $q > 2$ are not well supported by the data \cite{GW190521D}, therefore adding a parameter range to the model that will certainly increase the Occam factor and penalise the model. We perform a second run restricting $q \in [1,2]$. As expected we find a slightly increase evidence, so that $\log{{\cal{B}}}
   ^{\text{BBH},q\leq 2}_{{\text{BBH},q\leq 4}} = 0.6$. This slightly reduces the evidence for PHOCs to $\log{{\cal{B}}}
   ^{\text{PHOC},q = 1}_{{\text{BBH},q\leq 2}} = 0.2$ and $\log{{\cal{B}}}
   ^{\text{PHOC},q \neq 1}_{{\text{BBH},q\leq 2}} = 1.3$, but still favours this scenario.\\
   
    We further restrict the mass ratio to $q=1$ for the NRSur7dq4 model. The motivation for this is two-fold. First, this is the mass ratio of our primary PHOC model. Second, this is where the posterior distribution for the BBH model peaks \cite{GW190521D}, therefore leading to a stronger evidence. In fact, doing this we obtain $\log{{\cal{B}}}
   ^{\text{PHOC},q = 1}_{{\text{BBH},q = 1}} = - 0.3$ and $\log{{\cal{B}}}  
   ^{\text{PHOC},q \neq 1}_{{\text{BBH},q = 1}} = 0.8$, revealing a slight preference for the PHOC model, despite the strong intrinsic bias for BBH models introduced by our standard distance prior, as discussed in the main text and in the Supplementary Table \ref{tab:logb}.
\end{enumerate}

\begin{table}[h!]
\centering
\begin{center}
\begin{tabular}{lcc}
\rule{0pt}{3ex}
Waveform Model & $\log{\cal{B}}$ & $\log{\cal L}_{Max}$ \\
\hline
\rule{0pt}{3ex}%
Quasi-circular Binary Black Hole & 80.1 & 105.2\\
\rule{0pt}{3ex}%
Quasi-circular Non-precessing Binary Black Hole & 77.1 & 98.8\\
\rule{0pt}{3ex}%
Quasi-circular Binary Black Hole ($q \leq 2$) & 80.7 & 105.2 \\
\rule{0pt}{3ex}%
Quasi-circular Binary Black Hole ($q = 1$) & 81.2 & 105.2 \\
\rule{0pt}{3ex}%
Head-on Equal-mass Proca Stars &  80.9  & 106.7\\
\rule{0pt}{3ex}%
Head-on Unequal-mass Proca Stars &   82.0 & 106.5\\
\rule{0pt}{3ex}%
Head-on Binary Black Hole &  75.9 &  103.2 \\ 
\end{tabular}

\caption{\textbf{Impact of Occam penalty on model selection:} We show the same results as in Table I in the main text, adding restricted BBH models that may obtain larger evidences than the full ``vanilla'' one (first row) due to the reduction of the Occam penalty. In particular, we explore the effect of restricting to aligned spins and also to mass ratios $q\leq 2$ and $q=1$. None of these simplifications allows to make the models preferred wrt. our Proca star merger models.}
\label{tab:BBH_occam}
\end{center}
\end{table}

\subsection{Evidence for Head-on BBHs}
\begin{figure}
\begin{center}
\includegraphics[width=0.46\textwidth]{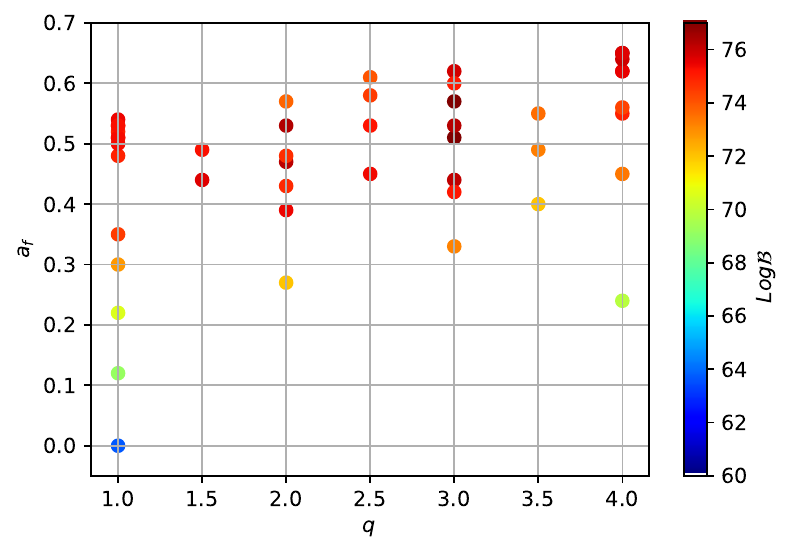}
\caption{\textbf{Numerical simulations for head-on BH mergers.} We label our simulations by the mass-ratio and the final-spin. The color code denotes the Bayesian evidence $\log{\cal{B}}$ obtained for each simulation, marginalised over the extrinsic parameters and the total mass of the source (Eq. 12).}
\label{fig:hoc_sims}
\end{center}
\end{figure}

For the HOC case, we did not explore the space spanned by the mass ratio and spins of the source in any systematic way. Instead, we performed simulations trying to maximise the Bayesian evidence (therefore populating the parameter space in a in-homogeneous way) until we determined it was not possible to keep increasing it. Fig. \ref{fig:hoc_sims} shows the Bayesian evidence marginalised over extrinsic parameters and total mass for each of our HOC simulations, as a function of the mass ratio and the final spin. The largest evidences are yielded by sources with mass ratio $q\in[2,3]$, which can lead to larger final spins than equal-mass systems. We find that increasing the mass ratio and the final spin further does not lead to an increase of the evidence, nor the log likelihood. For this reason our simulations only reach a mass ratio $q=4$.\\

Given our in-homogeneous family of HOCs, we cannot directly make use of to Eq. 8 to integrate $Z$ as there is no parameter on which our simulations span an uniform grid. Instead, we interpolate the marginalised Bayesian evidence across an uniform grid in final spin and mass-ratio, and compute the evidence for the whole model as:

\begin{equation}
\begin{aligned}
    \mathcal{Z}_{\text{HOC}} \approx \sum \mathcal{L}(d|(q,a_f),HOC)\, \pi(q,a_f)\, dq da_f,
\end{aligned}
\end{equation}
with
\begin{equation}
    \mathcal{L}(d|q,a_f;HOC) = \int {\cal{L}}(d|(q,a_f),\hat \theta,HOC)\, \pi{(\hat \theta)}\, d\hat \theta.
\end{equation}

Finally, given the evident lack of simulations below a final spin $a_f = 0.3$ we only include simulations with $a_f \geq 0.3$ in the above calculation.

\subsection{Constructing posterior distributions for HOC and PHOC}

Since the BBH model spans a continuous parameter space, we can trivially obtain posterior distributions on the different parameters marginalised over all other 14 parameters assuming given priors on these. However, our numerical simulations for HOCs and PHOCs span only a discrete set of mass ratios and spins.
For this reason, for these models, and for each value of the mass ratio and spins, we obtain a discrete set of posterior parameter distributions for the extrinsic parameters and the total mass, collectively denoted by $\hat \theta$.\\

To construct distributions marginalised over the intrinsic parameters of the simulations, we draw from each individual distribution for fixed mass and spins, a number of random samples proportional to the corresponding Bayes Factor. Note that since our PHOC simulations are uniformly distributed in the parameter $\omega/\mu_V$, we are intrinsically assuming an uniform prior on this parameter. In particular, for the parameter $\omega/\mu_V$ the distribution shown in Fig. 4 is given by $p(\omega/\mu_V) \propto \log\cal{B}(\omega/ \mu_V)$.\\

Given that our HOC simulations do not distribute in a rather arbitrary and non-uniform way across the parameter space, we cannot quote posterior parameter distributions under the assumption of any reasonable prior. For this reason, the estimates provided for the total mass and distance for HOC cases in the main text should be taken as rather ballpark numbers.\\ 

\textcolor{black}{Finally, we note that work like \cite{Gayahtri_ecc} chose to compute results based on single simulations for individual values of $\bar{\Theta}$, which in their case would denote individual combinations of mass-ratio and spins (and eccentricity). This approach, however, does not only lead to over-constrained estimates for the total mass and extrinsic parameters (distance, inclination, etc) but prevents an exact comparison of full source families that explicitly accounts for the size of the parameter space. In the case of \cite{Gayahtri_ecc}, it prevents to perform model selection between a non-eccentric and an eccentric BBH model. In contrast we perform full Bayesian selection between our quasi-circular BBH model and our other two models, namely our HOC and PHOC models, which is similar to the approaches in, e.g., \cite{GW190521D,Isobel_ecc}}.


\bibliography{IMBBH.bib}

\end{document}